\documentclass[11pt]{article}

\usepackage{makeidx}   
\usepackage{graphicx}  
\usepackage{subeqnar}  
\usepackage{multicol}  
\usepackage{cropmark} 
\usepackage{physprbb}  
\begin{document}

\vspace*{-1.8cm}
\begin{flushright}{\bf LAL 00-64}\\
{October 2000}\\
\end{flushright}
\vspace*{1cm}

\begin{center}
{\LARGE\bf Lower limit on the mass of the neutralino (LSP) at LEP with the 
ALEPH detector}
\end{center}

%
%

%
%
%
\vspace*{0.5cm}
\begin{center}
{\Large\bf Laurent Serin, on behalf of the ALEPH collaboration}
\end{center}

%
%
\begin{center}
{\it\large Laboratoire de l'Acc\'el\'erateur Lin\'eaire}\\
{IN2P3-CNRS et Universit\'e de Paris-Sud, BP 34, F-91898 Orsay Cedex, France} 
\end{center}


\begin{abstract}
The large amount of data accumulated at LEP2 by the ALEPH experiment 
has been used to search for supersymmetric particles. No signal has been found 
therefore limits have been determined. Within the 
Constrained Minimal Supersymmetric Standard Model, the constraints
from direct SUSY searches of charginos, sleptons and neutralinos,
are combined  to extract a lower limit on the mass of the neutralino 
considered to be  the Lightest  Supersymmetric particle. An improved limit is obtained
when the limit on the Higgs mass is included. Neutralino masses up to 38 GeV/c$^2$
are excluded at 95 \% confidence level.

\end{abstract}

\section{Introduction}
In many supersymmetric models conserving R-parity, the neutralino 
as Lightest Supersymmetric Particle, is considered 
to be a candidate for cold dark matter if its density does not over-close the 
Universe~\cite{Falk}. Experiments 
aiming at a direct detection of these Weakly Indirecting Massive Particles 
(WIMP) exist and are summarized in ~\cite{dark}. On the other hand, at 
accelerator experiment, a lower limit on the mass of the neutralino can be 
also extracted and results from the combination of direct SUSY searches
(charginos, sleptons,...).
 
First the SUSY  model used at LEP to obtain the excluded regions 
and the strategy to set a lower limit on the mass of the neutralino are 
discussed. Then the signal and background topologies
are explained. The exclusion and the limit on the neutralino mass are shown 
in section 3, including a discussion of the main sources of theoretical 
uncertainties. Finally, the impact of Higgs search is presented.

\section{Definition of SUSY framework and strategy}

Most of the results are interpreted at LEP in 
the Constrained Minimal Supersymmetric Standard Model (CMSSM)
with R-parity conservation and 
with the assumption of GUT mass unification.  In this model a few parameters are 
enough to describe the properties of SUSY particles :
\begin{itemize}
\item m$_0$ : the common sfermion mass at the GUT scale
\item m$_{1/2}$ : the common gaugino mass at the GUT scale
\item A$_0$ : the universal trilinear coupling
\item tan$\beta$ : the ratio of the two higgs doublet vacuum expectation values
\item $\mu$ : the higgsino mass parameter
\item m$_A$ : the CP-odd Higgs boson mass, relevant for the Higgs sector
\end{itemize}
 
The corresponding parameters at the electroweak scale
are obtained by solving the Renormalization Group Equations. 
In practice M$_2$ is used as a free parameter instead of m$_{1/2}$.
From m$_0$, M$_2$ and tan$\beta$ the slepton masses are derived, 
with the typical hierarchy : $ m_{\tilde{q}} \gg m_{\tilde{\ell}_L} \sim m_{\tilde{\nu}} 
> m_{\tilde{\ell}_{R}}$.
As an example the slepton right mass can be written as : 
\begin{equation}
m^2_{\tilde{\ell}_R}= {{\bf m_0}^2+0.22 {\bf M_2}^2 -\sin^2\theta_\mathrm{W} M_Z^2{{\bf {cos2 {\bf\beta} }}}}
\end{equation}

The chargino and neutralino masses and couplings are 
fully specified by M$_2$, $\mu$, tan$\beta$.  

If mixing in the scalar sector is not considered, the set of free parameters
is reduced  to :  M$_2$, $\mu$, tan$\beta$, m$_0$.  A scan over this 4-parameter space
is performed to search for the lowest allowed neutralino mass. 
Two regions, with distinct characteristics depending on the scalar mass m$_0$, exist. 
For large scalar masses, chargino and neutralino 
limits  are used. For low scalar masses, sleptons results must be included.
Additional regions are excluded by Higgs constraints, especially for low tan$\beta$.
In order to be conservative, neutralino decays involving Higgs bosons are usually inhibited. 
\section{Experimental context}
Since 1995, the LEP2 $e^+ e^-$ collider has regularly increased 
its center of mass energy beyond the Z resonance. Table~\ref{Tab1} shows the luminosity 
accumulated by ALEPH up to 1999.  The results presented in this paper 
use mainly the data accumulated at and above 189 GeV.

\begin{table}[h]
\caption{Luminosity accumulated by the ALEPH experiment}
\begin{center}
\renewcommand{\arraystretch}{1.4}
\setlength\tabcolsep{5pt}
\begin{tabular}{lll}
\hline\noalign{\smallskip}
Year & $\sqrt{s} $ & {$\cal{ L}$} \\
\noalign{\smallskip}
\hline
\noalign{\smallskip}
1995 & 130-136 GeV & $\simeq 5.7 $ pb$^{-1}$ \\
1996 & 161-172 GeV & $\simeq 21 $ pb$^{-1}$ \\
1997 & 181-184 GeV & $\simeq 57 $ pb$^{-1}$ \\
1998 & 189 GeV     & $\simeq 174 $ pb$^{-1}$ \\
1999 & 192-202 GeV & $\simeq 237 $ pb$^{-1}$ \\
\hline
\end{tabular}
\end{center}
\label{Tab1}
\end{table}

The main characteristic of SUSY signals is the presence of 
neutralinos in the final state escaping detection, leading to 
acoplanar jets, acoplanar leptons or mixed jet/lepton topology. 
These events have a large 
missing mass and missing transverse momentum which depends strongly on the 
mass difference ($\Delta M$) between the produced particle (chargino, slepton 
or heavy neutralino) and the neutralino (LSP), one of the decay products. 
Specific analyses have been designed depending on the mass difference and 
on the final state ~\cite{aleph1} . 

\begin{figure}[h]
\begin{center}
\includegraphics[width=.45\textwidth]{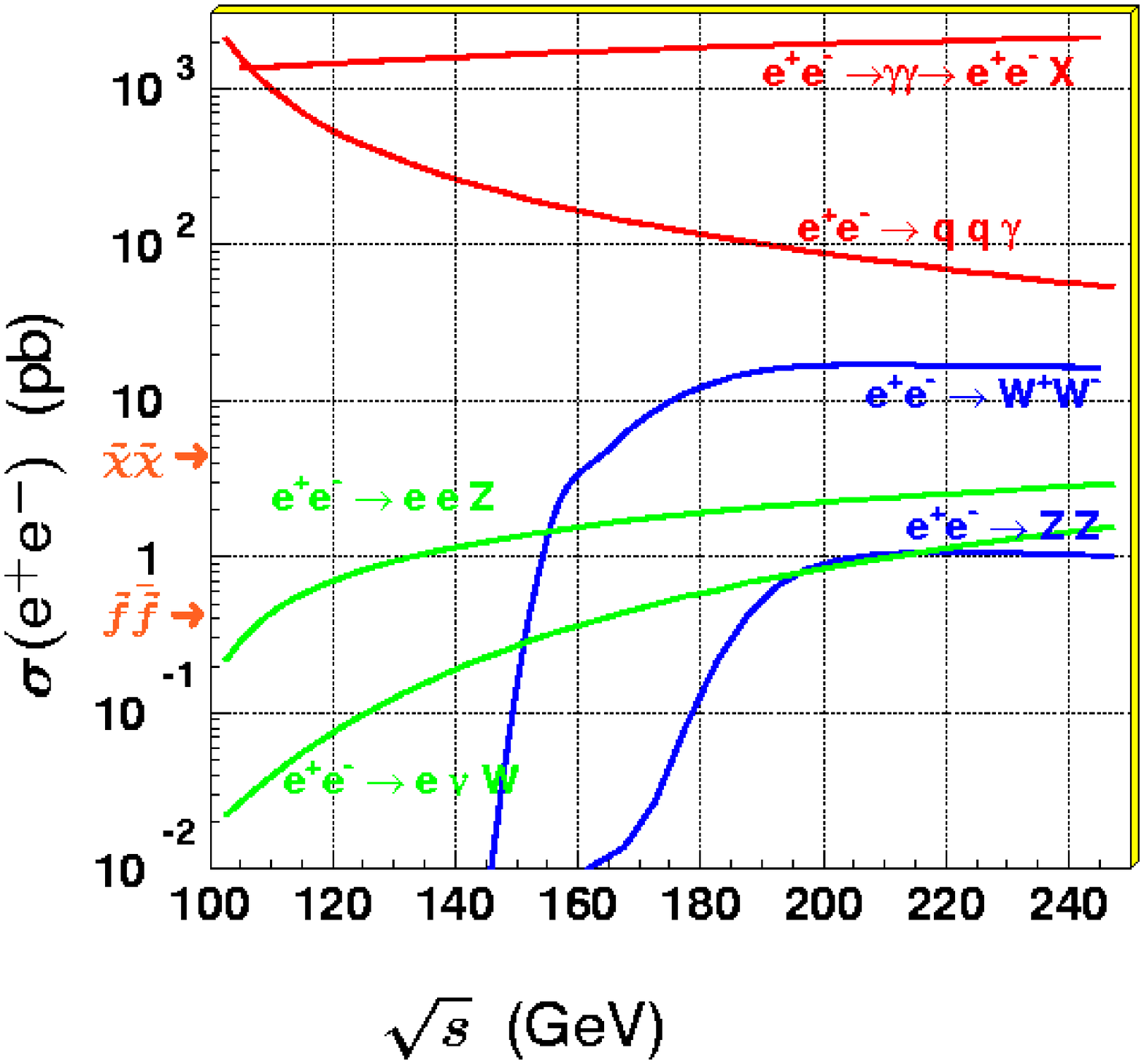}
\end{center}
\caption[]{Cross section of Standard Model backgrounds as function of the center of mass
energy}
\label{Fig1}
\end{figure}

The cross section of the main Standard Model backgrounds is presented in 
Figure \ref{Fig1} : 
\begin{itemize}
\item For small $\Delta M$, the $\gamma \gamma$ background is typically 
three orders of magnitude larger than expected signal, but cuts on missing mass or missing 
transverse momentum reduce this background to a very low level.
\item For large  $\Delta M$, the QCD background can be easily rejected and 
the dominant background (WW, We$\nu$ and ZZ) is almost irreducible for some final states.
\end{itemize}

\section{Excluded region and limit on the neutralino mass}
\subsection{Sleptons}
The sleptons are produced by pair in the s-channel via $\gamma, Z$, 
and in the t-channel via neutralino exchange, leading to an increase of the 
cross section for selectron production.  The final state consists
of two leptons and two neutralinos (the efficiency for cascade 
decays is assumed to be zero). No evidence of a signal was observed, e.g. in the 
1999 data, 42 events were observed while 39 were expected from background \cite{winter}.
The main 
background comes from leptonic decays of WW events (the WW background is subtracted
in all analyses involving at least one lepton).
Figure \ref{Fig2} shows the limits obtained in the selectron-neutralino plane. 
Typically selectron masses up to 95 GeV/c$^2$ are excluded. As shown in equation 1, at fixed tan$\beta$
and for small scalar masses m$_0$, this limit can be translated directly on a limit on M$_2$.

\subsection{Charginos}
\begin{figure}
\begin{center}
\includegraphics[width=8cm]{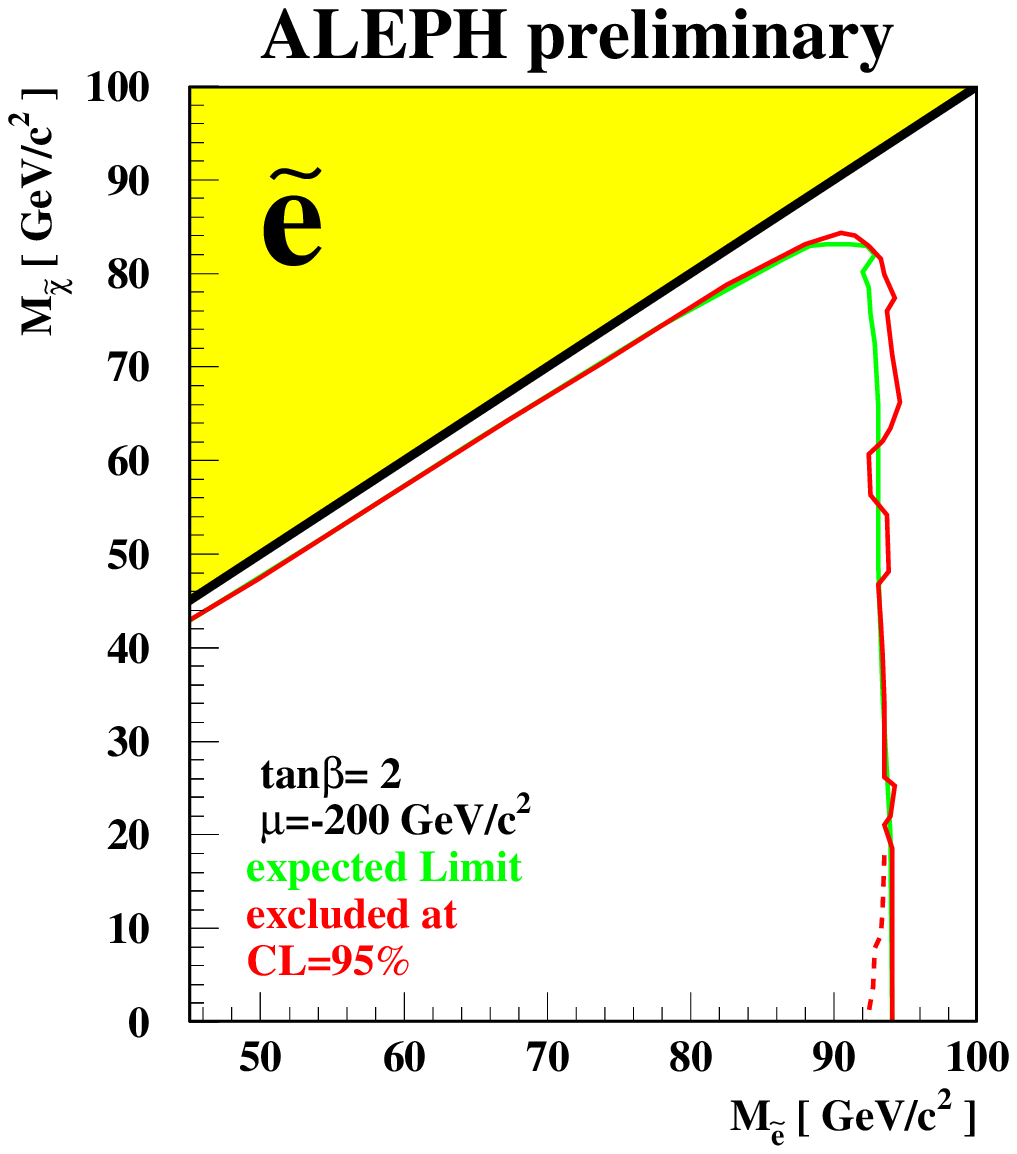}
\includegraphics[width=8cm]{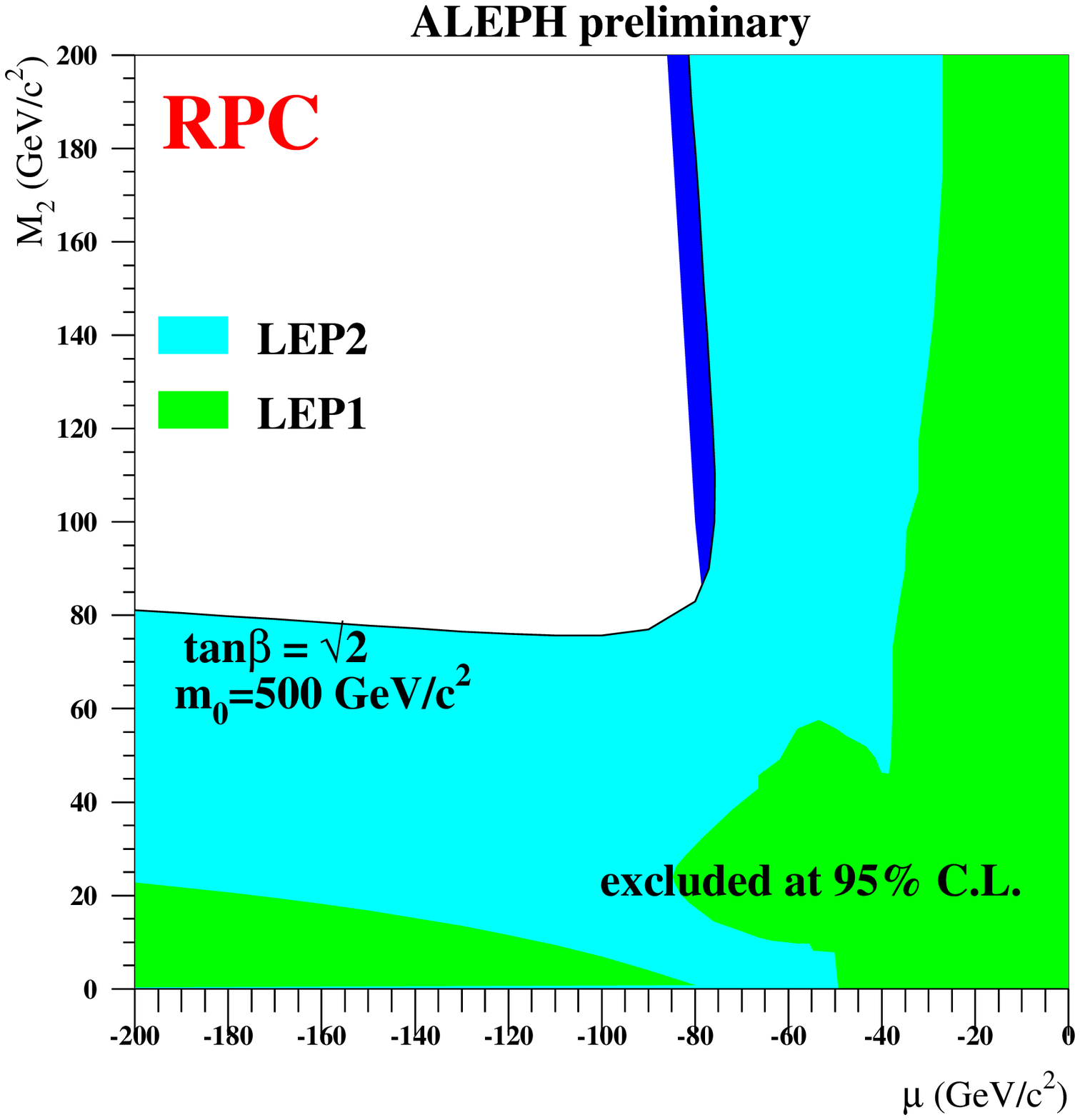}
\end{center}
\vspace*{-0.5cm}
\caption[]{ALEPH limit from selectron search in the selectron-neutralino mass plane. The yellow
part is forbidden by theory. (left) and 
from neutralino and chargino searches for large scalar masses in the (M$_2$,$\mu$) plane.
The dark blue area shows the improvement of the limit with neutralino with respect
to chargino exclusion alone (right).}
\label{Fig2}

\begin{center}
\includegraphics[width=.49\textwidth]{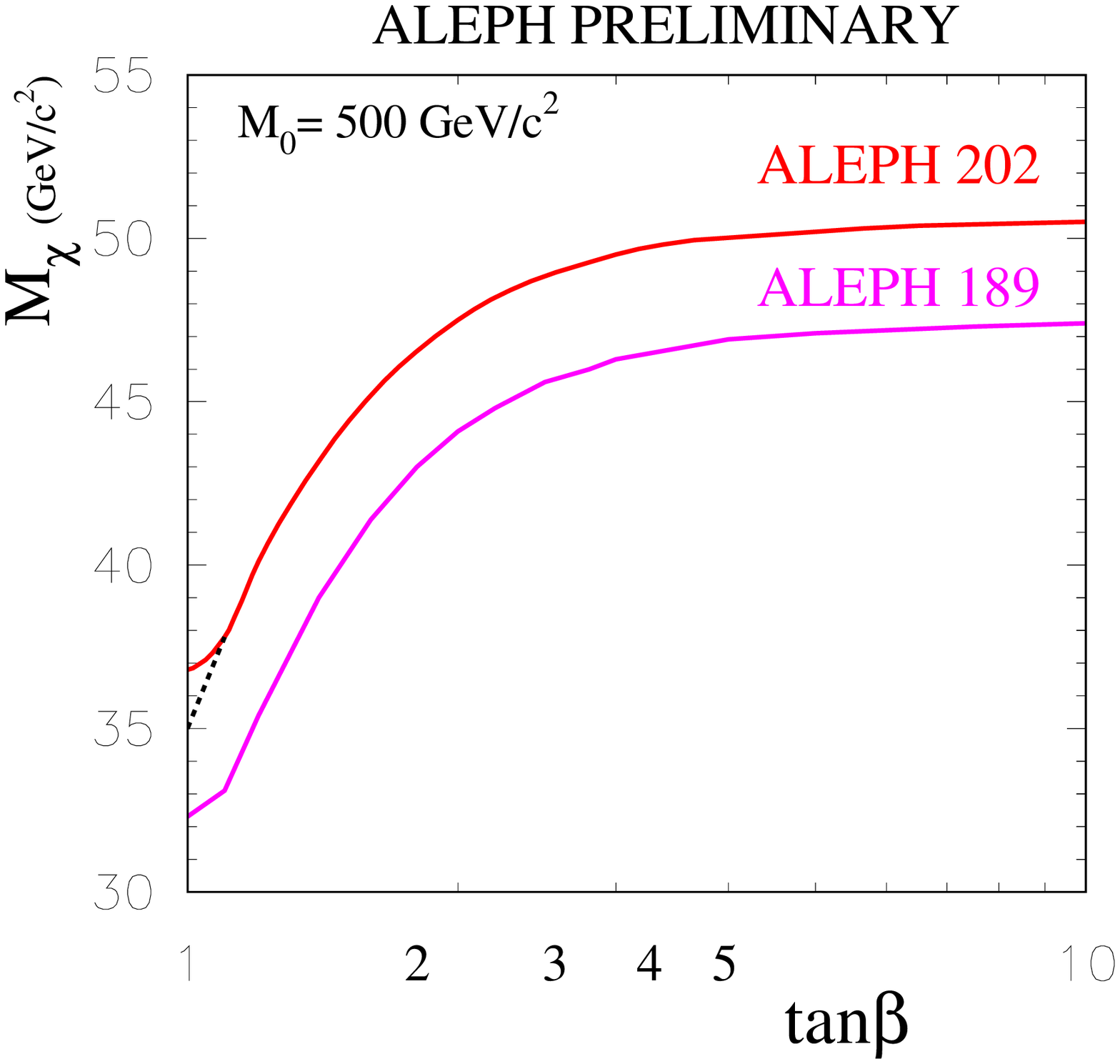}
\includegraphics[width=.49\textwidth]{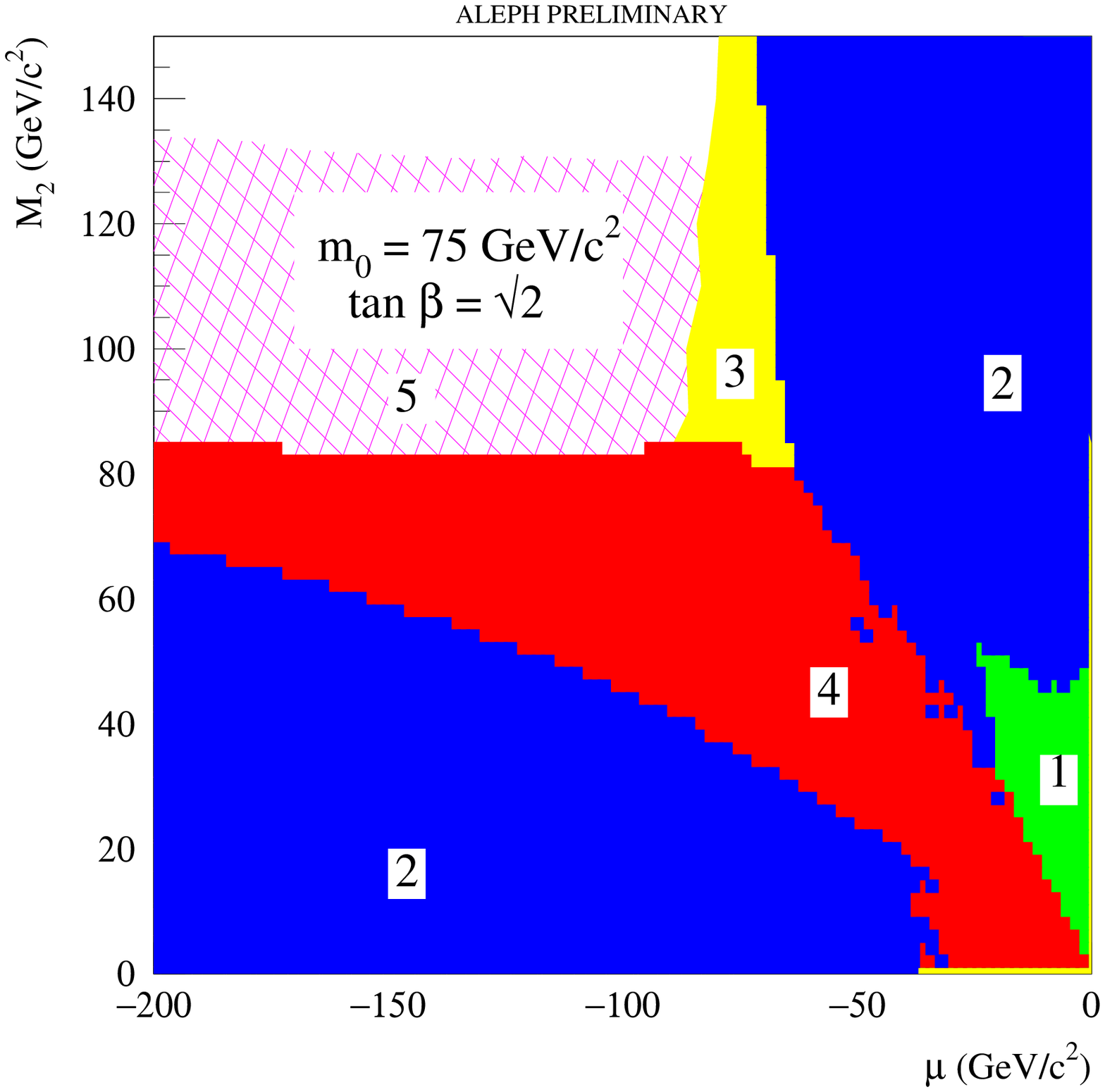}
\end{center}
\vspace*{-1cm}
\caption[]{Limit on the neutralino mass for m$_0 = 500 $ GeV/c$^2$ versus tan$\beta$. 
The dashed line shows the limit obtained with charginos only (left). 
Illustration of the interplay among the various searches for low 
scalar masses with the 183 GeV data : LEP1 (1), chargino (2), neutralino (3), slepton (4) and Higgs (5)
The corridor can be observed  between the chargino exclusion in the higgsino and gaugino 
regions  (right).}
\label{Fig4}
\end{figure}

Charginos are produced in the s-channel wia $\gamma, Z$ 
and in the t-channel with sneutrino exchange. This two diagrams interfere 
destructively for low scalar masses resulting in a reduction of the  
cross section. The decay proceeds via a three-body decay (W$^*$) or a two-body decay 
(sneutrino-lepton for instance), 
leading to final states of 4-jets, 2-jets-lepton, and 2-leptons. A map of the efficiency 
and background expectation has been computed as a function of $\Delta M$ and 
the leptonic branching ratio. For heavy (light) sfermions, 9 (24) candidates are observed in 1999 data, 
while the expectation from background is 12.7 (33.9). The chargino exclusion for large scalar 
masses, shown in Figure~\ref{Fig2} reaches the kinematic limit. 
For low scalar masses, when the chargino and sneutrino are almost mass-degenerate (a few GeV), 
the final state is practically  invisible 
as the lepton energy, from the chargino decay, is too soft, leading to a non excluded corridor
which is displayed on figure~\ref{Fig4}.

\subsection{Neutralinos}
In contrast to  charginos, the s-channel and the t-channel (via slepton exchange), 
show a constructive interference, thus an increase of the cross section for low
scalar masses. All neutralino production channels 
($\chi_1^0 \chi_2^0, \chi_1^0 \chi_3^0 \chi_1^0 \chi_4^0$...)
are taken into account . Final states considered are acoplanar jets and acoplanar leptons for low m$_0$.
For heavy (light) sfermions, 5 (78) candidates are observed, while 3.1 (87.7) are expected 
from background. For large scalar masses, the neutralinos are useful in the higgsino region 
($|  \mu |  > 100 $) for low tan$\beta$ and negative $\mu$ as shown 
in figure~\ref{Fig2} as they exclude region above the chargino kinematic limit.

\subsection{Limit on the neutralino mass}

For large scalar masses the chargino and neutralino exclusions have been combined
and Figure~\ref{Fig4} shows the limit on the neutralino mass as a function of 
tan$\beta$. The limit, given by neutralinos with $\chi_4^0 \chi_2^0$ production for 
tan$\beta = 1$, is 
$m_{\chi}> $ 37 GeV/c$^2$ at 95 \% CL for m$_0 >$  500 GeV/c$^2$.

To extract the absolute limit obtained for any m$_0$, first the chargino and slepton searches  
are combined via parametrisations of efficiency and maps of background. 
Then for each (m$_0$, tan$\beta$), for all the points not yet excluded with a neutralino mass
$<$ 38 GeV/c$^2$,  neutralino events are simulated, reconstructed and the analyses applied
to determine if the point is excluded.  The combined limit is : 
\begin{center}
$m_{\chi} > $  35 GeV/c$^2$ at 95 \% CL for any m$_0$ and tan$\beta$
\end{center}
\subsection{Uncertainties on the LSP mass limit}

A few caveats concerning the robustness of the LSP mass
limit should be mentioned  :
\begin{itemize}
\item The dependence of the limit on the mixing parameters, 
which induce an enhancement of chargino and neutralino
decays to taus, must be studied. 
\item Radiative corrections to chargino and neutralino masses 
decrease the limit by about 1 GeV in the region where the limit
is found. 
\item Higher order corrections to the one-loop GUT relation
between M$_1$ and M$_2$ introduce a variation of a few $\%$ 
on the LSP mass limit.
\end{itemize}

The  typical uncertainty on the LSP mass limit is therefore about 2 GeV. 

\section{Higgs constraints on the neutralino mass limit}

\begin{figure}[htb]
\begin{center}
\includegraphics[width=.45\textwidth]{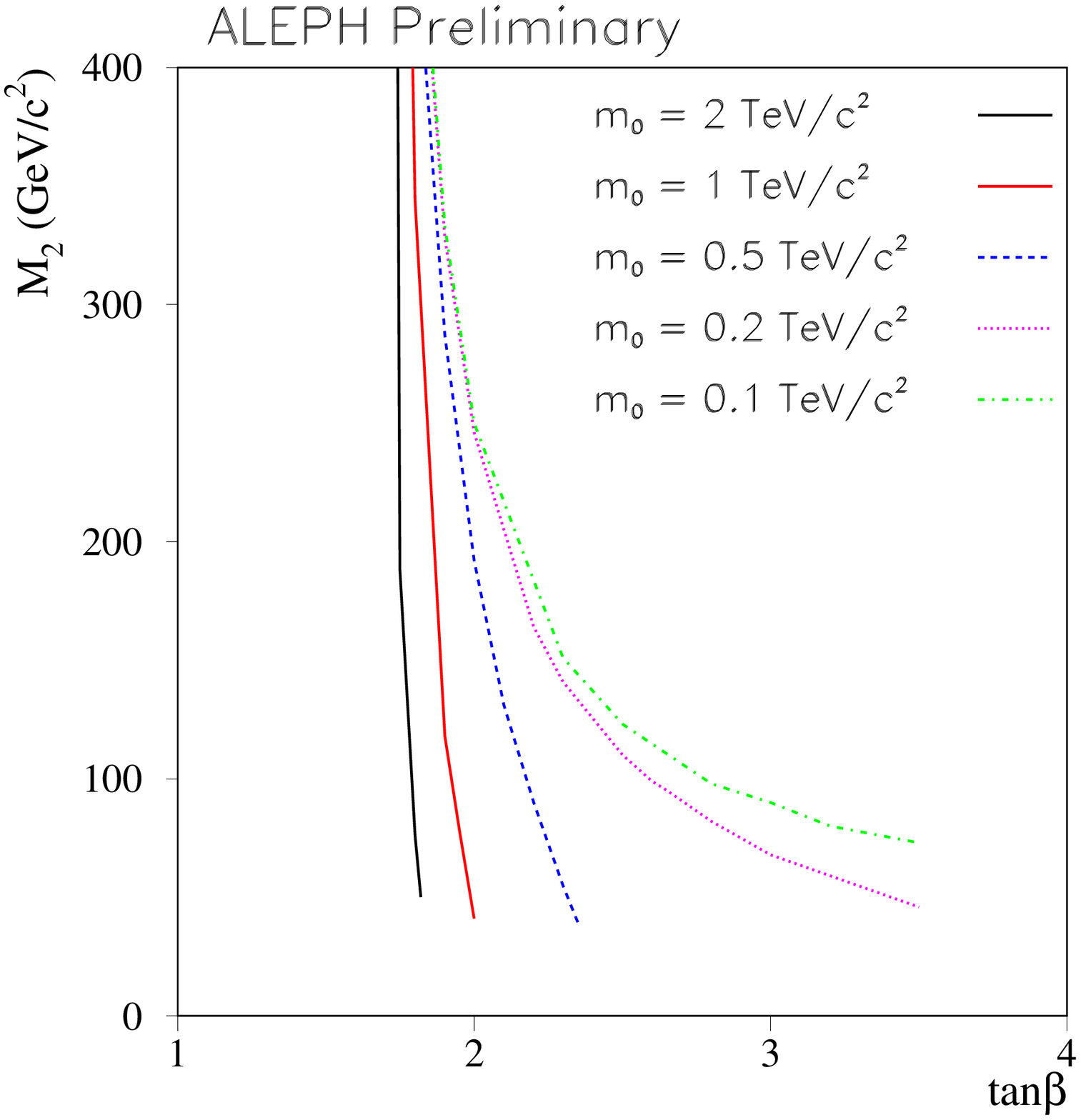}
\includegraphics[width=.45\textwidth]{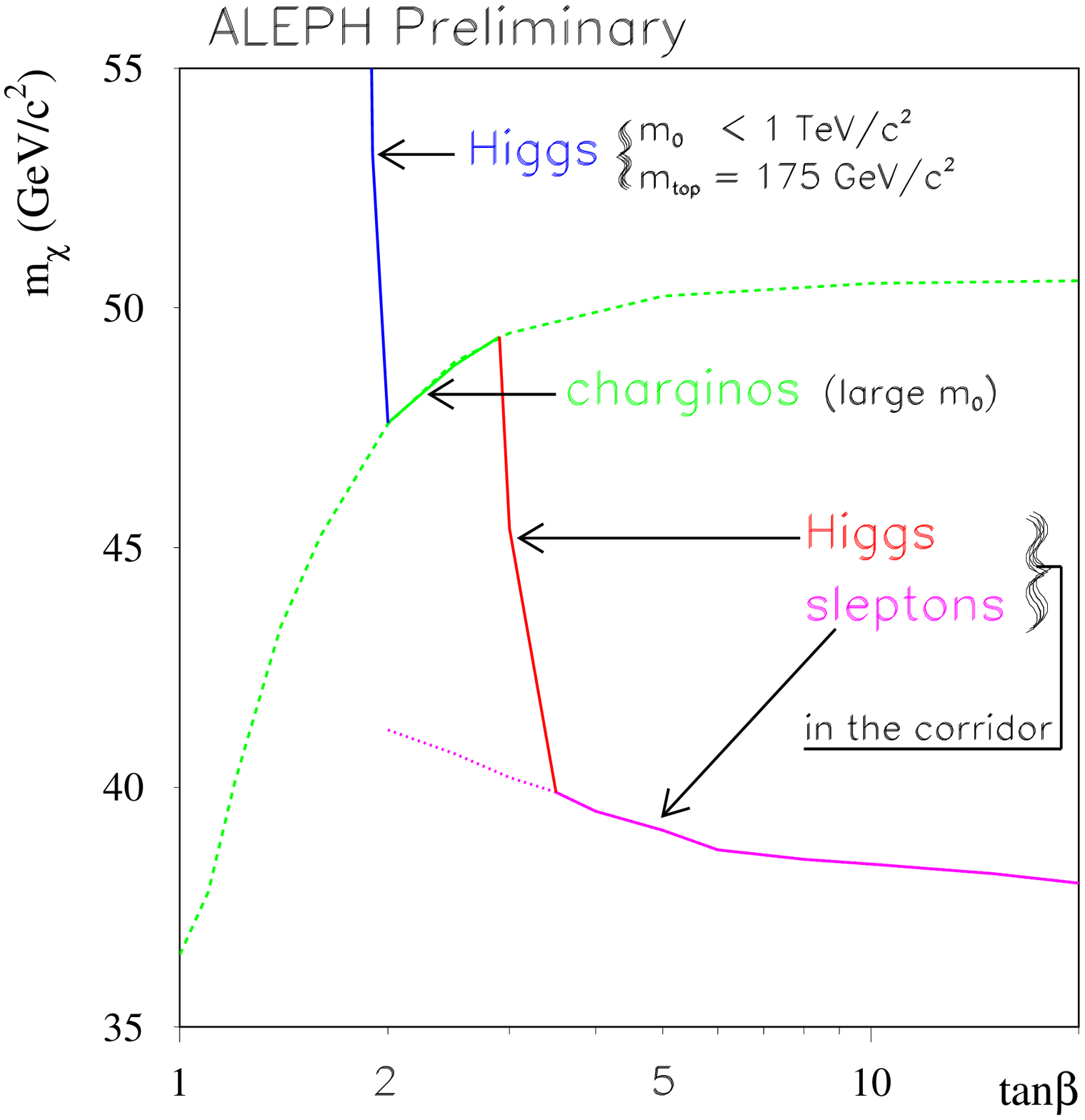}

\end{center}
\caption[]{Higgs constraints in the (M$_2$,tan$\beta$) plane (left), and limit on the 
neutralino mass combining chargino, neutralino, slepton and Higgs contraints versus
tan$\beta$ (right)}
 
\label{Fig5}
\end{figure}

At tree level, the mass of the lightest supersymmetric Higgs, m$_h$ depends only on m$_A$ and
 tan$\beta$, and  is bounded by  $m_Z | cos 2 \beta |$. A limit on m$_h$ therefore gives directly a limit 
on tan$\beta$ as soon as m$_A$ is fixed. As the radiative corrections to the Higgs 
mass introduce a dependence on the stop mass, i.e.  the mixing parameter A$_t$, the picture 
has to be refined. Fortunately large stop mixing, which means an increase of the Higgs mass,
induces also a small stop mass. Using the Higgs constraints for small mixing, and the stop search
for large mixing, the exclusion obtained previously in the (M$_2$, $\mu$) plane can be substantially improved 
for low m$_0$ and tan$\beta$. A scan over all possible values on m$_A$ and A$_t$ within the physical ranges
is performed to obtain this limit.  As an illustrative example, the improvement is shown in figure~\ref{Fig4}
at 183 GeV. To update the results for a higher center of mass energy, the stops constraints are no longer
useful and the (tan$\beta$, m$_0$, M$_2$, $\mu$) plane is scanned with the following procedure
\cite{higgs} :  
\begin{itemize}
\item It has been shown that the dependence of the limit with $\mu$ is negligible, the value 
has been fixed to $-$100 GeV/c$^2$ where m$_h$ is maximal.
\item m$_A$ has been fixed to 2 TeV/c$^2$ to maximize m$_h$. A scan over A$_t$ is performed 
to determine the highest value  
of $m_h$ in each point, which is compared to the ALEPH Higgs mass limit on m$_h$  (107.7 GeV/c$^2$)
(m$_t$ = 175 GeV/c$^2$ is used).
\item The Higgs mass limit, if the radiative corrections are small,  can be 
translated into a limit on tan$\beta$. In the other case, this limit can be translated into a limit on the 
stop mass, and therefore into a limit on $M_2$ (assuming m$_0$ is not too large). 
\end{itemize}
Figure~\ref{Fig5} summarizes the exclusion obtained in the (M$_2$,tan$\beta$) plane from the Higgs limit. 
Any value of tan$\beta < 1.7$ is excluded.   When increasing tan$\beta$ up to about 3,  
a lower limit on M$_2$ is obtained for small scalar masses (typically $<$  200 GeV/c$^2$). 
The result of the combination of the Higgs constraint with the previous limit on the neutralino 
mass is shown in Figure~\ref{Fig5}. Low tan$\beta$ are excluded by Higgs constraints and supersedes 
the chargino/neutralino limit. However this limit is very sensitive, through the radiative corrections,  
to the top mass. If m$_t = 180 $ GeV/c$^2$, this exclusion do not hold, but a 40 GeV/c$^2$ neutralino mass 
is always excluded when tan$\beta < 3 $. If furthermore m$_0$ is taken to be 2 TeV/c$^2$, no 
more limit from Higgs can be used.  Then the limit is given by the charginos exclusion at large scalar masses in a tiny
region of tan$\beta$. Finally increasing with tan$\beta$, the neutralino limit is found in the corridor where
charginos and neutralinos can not help. For tan$\beta > 3$, only the slepton search 
contributes and the absolute limit is~:
\begin{center}
$m_{\chi} > $ 38 GeV/c$^2$ at 95 \% CL for any m$_0$ and tan$\beta$
\end{center}

\section{Conclusion}

Using the large amount of data accumulated at LEP2 by ALEPH , the combination of direct searches 
of SUSY particles can be used to extract a lower limit on the mass of the neutralino, as
as the Lightest Supersymmetric particle. A lower limit of 35 GeV/c$^2$ on the neutralino mass is obtained
for any scalar mass m$_0$ and tan$\beta$.  Moreover, a limit of 38 GeV/c$^2$
 is obtained including the Higgs contraints.  At the end of LEP operation, a value of at least 40 GeV/c$^2$ should be
reached if no signal is observed. 
\newpage

%


\begin{thebibliography}{8.}
\bibitem{dark} More details on the existing and future experiments in this 
 domain can be found in these proceedings
\end{thebibliography}
\end{document}